\documentstyle[prl,aps,preprint,epsf,graphicx]{revtex}
\begin{document}
\draft
\bibliographystyle{prsty} 
\tightenlines

\title{ Geometric Resonance in Modulated Quantum Hall Systems Near
$ \nu = 1/2$} 
\author{Nataliya A. Zimbovskaya$^{*}$ and Joseph L. Birman$ ^{**}$}
\address{Department of Physics, The City College of CUNY, New York,
NY, 10031, USA} 
 
\date{\today} 
\maketitle

\begin{abstract}
 We propose a theory for the new effects recently observed by Willett et
al [1] in the magnetoresistance of a weakly modulated two dimensional
electron gas near filling factor 1/2. Minima in transverse
magnetoresistance and maxima in longitudinal magnetoresistance at the same
magnetic field producing the new resonance structure are reported. The
structure occurs due to geometric resonance of the composite fermion
cyclotron orbits with the modulation period of the effective magnetic
field $\, B_{eff} \,$ due to the applied density modulation. The
transverse minimum occurs due to the inhomogeneity in the field $
\,B_{eff} \,$ in the presence of density modulations, whereas the
longitudinal maximum can arise due to a shape-effect (distortion) of the
composite fermion Fermi surface (CF--FS). Thus the minima and maxima
reflect different physical mechanisms. 
              \end{abstract}

PACS numbers 71.10 Pm, 73.40 Hm, 73.20 Dx {}
\vspace{4mm}

\newpage
The present work is motivated by the new experimental results of dc
transport experiments of Willett et al [1] in a two-dimensional electron
gas (2DEG) in the fractional quantum Hall regime near half filling of the
lowest Landau level $ (\nu = 1/2) .$ The 2DEG was modulated with a density
modulation of small period applied in one direction. A new resonance
structure produced by the modulation was observed in the magnetoresistance
of the 2DEG. The resonance structure was superimposed on a minimum which
occurs at about $ \nu = 1/2 $ in the magnetic field dependence of that dc
resistivity component for a current driven {\bf across} the modulation
lines $ (\rho_\perp), $ and on a maximum in the resistivity corresponding
to a current driven {\bf along} the modulation lines $ (\rho_{||})$.  Such
a maximum in the magnetic field dependence of the dc resistivity of a
modulated 2DEG was never observed before.

Here we develop a semiquantitative theory of these results [1]. Our work
is based on the theory of the quantum Hall system at and near $ \nu = 1/2
$ proposed by Halperin, Lee and Read (HLR)  [2,3], which corresponds to
the physical picture of the electrons decorated by attached quantum flux
tubes. These are the relevant quasiparticles of the system -- so called
composite fermions (CF). The CFs are charged spinless fermionic
quasiparticles which move in the reduced effective magnetic field $
B_{eff} = B - 4 \pi \hbar c n /e \;  (n $ is the electron density) [4]. At
$ \nu = 1/2 $ the CFs form a Fermi sea and exhibit a FS. Within the HLR
theory the CF--FS can be taken as a circle in quasimomentum space. Its
radius $ p_F $ equals $ \sqrt{4 \pi n \hbar^2}. $

The density modulation influences the CF system in two ways: through the
direct effect of the modulating potential which can deform the CF--FS [5],
and also through the effect of an additional inhomogeneous magnetic field
$ \Delta B \bf (r) $ proportional to the density modulation $ \Delta n
{\bf (r)} $ [6]. To analyse the effect systematically we have to solve the
Boltzmann transport equation for the CF distribution function in the
presence of a spatially inhomogeneous disturbance due to the density
modulation in a similar fashion to the work of Ref.[7] for the 2DEG in a
low magnetic field. When, however, the CFs mean free path $ l $ is larger
than the radius of their cyclotron orbit at the effective magnetic field $
R $ and the period of modulation $ \lambda, $ we can obtain the desired
response functions using simplifications based on the work of Beenakker
[8] and Gerhardts [9]. 
The result of our work is that the deformation of the CF--FS affects
principally the longitudinal response, while the additional field $
\Delta B \bf (r) $ principally affects the transverse response.

We start from the Lorentz force equations describing the CF motion along
the orbit:
         \begin{equation}
\frac{d p_x}{d t} = - \frac{|e|}{c} B ({\bf r}) v_y; 
\qquad \qquad
\frac{d p_y}{d t} =  \frac{|e|}{c} B ({\bf r}) v_x; 
                               \end{equation} 
 where $ p_{x,y} $ and $v_{x,y} $ are the components of the CF
quasimomentum and velocity;\\ $ B{\bf (r)} = B_{eff} + \Delta B ({\bf r}) 
\equiv B_{eff} - 4 \pi \hbar c \Delta n {\bf (r)} /e.$

We will consider first a single-harmonic sinusoidal density modulation of
period $ \lambda = 2 \pi / g $ along the $ ''y'' $ direction: $ \Delta n
{\bf (r)} \equiv \Delta n (y) = \Delta n \sin (g y) .$ We assume that the
correction term $ \Delta B \bf (r) $ is small compared to $ B_{eff}. $
Under this assumption we can write the CF velocity $ \bf v $ in the form $
{\bf v = v}_0 + \delta \bf v, $ where $ {\bf v}_0 $ is the uniform-field
velocity and the correction $ \delta \bf v $ arises due to the
inhomogeneity of the magnetic field. For a circular CF--FS we have:  $
v_{x0} = v_F \cos \Omega t ; \; v_{y0} = v_F \sin \Omega t; \Delta n (y)
\approx \Delta n \sin (gY - g R \cos \Omega t) ,$ where $ v_F $ is the
CF's Fermi velocity, and $ Y $ is the $ ''y''$ coordinate of the guiding
center.  Substituting these expressions for $ \bf v $ and $ \Delta n (y) $
into Eqs.(1) and keeping only first-order terms we obtain: 

  $$
\frac{d (\delta v_x)}{dt} = - \Omega \delta v_y - \frac{\Delta B}{B_{eff}}
\Omega v_F \sin \Omega t \sin (g Y - g R \cos \Omega t);
             $$ 
        \begin{equation}
\frac{d (\delta v_y)}{dt} =  \Omega \delta v_x + \frac{\Delta B}{B_{eff}}
\Omega v_F \cos \Omega t \sin (g Y - g R \cos \Omega t).
                               \end{equation} 

We  remark here that in the presence of the density modulation the
CF's Fermi velocity gets a correction due to the modulation. To
evaluate this correction we calculate the average of $ \Delta n (y)$ over
the cyclotron orbit. Expanding the functions $ \cos (gR \cos \Omega t) $
and $ \sin (gR \cos \Omega t) $ in Bessel functions we arrive at the
following result for the averaged correction to the inhomogeneous density
modulation: 
         \begin{equation}
<\Delta n(y)> \equiv \frac{\Delta n}{2 \pi} \int \limits_0^{2 \pi} \sin
(g Y - gR \cos \psi) d \psi = \Delta n \sin (g Y) J_0 (g R).
                               \end{equation} 
 Here $ \psi = \Omega t. $

The result (3) gives a spatially inhomogeneous correction to the chemical
potential of the CFs and to their Fermi velocity:  $\displaystyle{ \tilde
v_F = v_F \sqrt{1 + \frac{\Delta n}{n} \sin (gY) J_0 (gR )}}.$ However we
can neglect the difference between $ v_F $ and $ \tilde v_F $ in the
equations (2) because it gives corrections which are an order of magnitude
smaller than those which are kept.  It is natural to suppose that to the
first order in the modulating field the corrections $ \delta v_x $ and $
\delta v_y $ are periodic over the unperturbed cyclotron orbit. This
assumption is equivalent to that used in Ref.[9]. Under this assumption we
can calculate averages of Eqs.(2) over the cyclotron orbit.  This gives us
the following expressions for the components of the velocity of the
guiding center drift $ V_x $ and $ V_y $ defined below: 
 $$  
V_x (Y) = <\delta v_x> = - \frac{v_F}{2 \pi} \frac{\Delta B}{B_{eff}} 
\int \limits_0^{2 \pi} \cos \psi \sin (g Y - g R \cos \psi) d \psi =
v_F \frac{\Delta B}{B_{eff}} \cos g Y J_1 (g R);
                $$
         \begin{equation}
V_y (Y) = <\delta v_y> = - \frac{v_F}{2 \pi} \frac{\Delta B}{B_{eff}} 
\int \limits_0^{2 \pi} \sin \psi \sin (g Y - g R \cos \psi) d \psi = 0.
                               \end{equation} 

To evaluate semiquantitatively the CF conductivity we assume that the $ x 
$ component of the CF velocity can be written in the form $ v_x (Y) =
v_{x 0} + V_x (Y). $ We also assume that the cyclotron frequency $
\Omega $ can be replaced by the quantity $ \Omega (Y) = \Omega + <\Delta 
\Omega (y)> $ where $ <\Delta \Omega (y) > $ is the correction to the
cyclotron frequency arising due to the inhomogeneity of the effective
magnetic field averaged over the cyclotron orbit:
     $$
\Omega (Y) = \Omega \left \{1 + \frac{\Delta B }{B_{eff}} \frac{1}{2\pi}
\int \limits_0^{2 \pi} \sin (gY - gR \cos \psi) d \psi \right \} 
      = $$
         \begin{equation}
= \Omega
\left \{1 + \frac{\Delta B}{B_{eff}} \sin (gY) J_0 (gR) \right \}.
                               \end{equation} 
 We showed before [5] that in the semiquantitative analysis of the
magnetotransport in the modulated the 2DEG we can use the following
approximation for the CF conductivity: 
         \begin{equation}
\sigma_{\alpha \beta}^{cf} \approx \frac{g}{2 \pi} \int \limits_{-\pi
/g}^{\pi /g} \sigma_{\alpha \beta}^{cf} (Y) d Y,
                               \end{equation} 
 where
         \begin{equation}
\sigma_{\alpha \beta}^{cf} (Y) = \frac{e^2 m_c \tau}{2 \pi \hbar^2} 
\sum \limits_k \frac{v_{k \beta} (Y) v_{- k \beta} (Y)}{1 + i k \Omega
(Y) \tau}.
                               \end{equation} 
 Here $ m_c $ is the CF cyclotron mass; $ \tau $ is the relaxation time; $
v_{k \beta} (Y) $ are the Fourier transforms for the CF velocity
components:  $ \displaystyle{ v_{kx} = \frac{\tilde v_F}{2} (\delta_{k,1}
+ \delta_{ k,-1}) + V_x (Y) \delta_{k,0}; \;  v_{ky} = \frac{i \tilde
v_F}{2}} (\delta_{k,1} - \delta_{ k,-1}) .$

Keeping  terms of the order of $ (\Delta B /B_{eff})^2 $ or larger we
obtain the following approximations for the CF conductivity components $
\displaystyle{\left (\frac{\Delta B}{B_{eff}} \Omega \tau =
\frac{\Delta n}{n} \frac{p_F}{\hbar} l = \frac{\Delta n}{n} k_F l
\right ): }$
          \begin{equation}        
\sigma_{xx}^{cf} \approx \frac{\sigma_0}{1 + (\Omega \tau)^2} \left \{1 +
\frac{3}{2} \left (\frac{\Delta n}{n} k_F l \right )^2 \frac{J_0^2
(gR)}{1 + (\Omega \tau)^2} \right \} +
\sigma_0 \left (\frac{\Delta B}{B_{eff}} \right )^2 J_1^2 (g R);
                               \end{equation} 
         \begin{equation}
\sigma_{yy}^{cf} \approx \frac{\sigma_0}{1 + (\Omega \tau)^2} 
\left \{1 + \frac{3}{2} \left (\frac{\Delta n}{n} k_F l \right )^2 
\frac{J_0^2 (gR)}{1 + (\Omega \tau)^2} \right \};
                               \end{equation} 
         \begin{equation}
\sigma_{xy}^{cf} = - \sigma_{yx}^{cf}
\approx \frac{\sigma_0 \Omega \tau}{1 + (\Omega \tau)^2} 
\left \{1 + \frac{1}{2} \left (\frac{\Delta n}{n} k_F l \right )^2 
\frac{J_0^2 (gR)}{1 + (\Omega \tau)^2} \right \}.
                               \end{equation} 
 where $ \sigma_0 = n e^2 l / p_F $ is the CF conductivity in a
homogeneous magnetic field.

The last term in the expression for $ \sigma_{xx}^{cf} $ describes the
contribution from CFs diffusing along the $ ''x'' $ direction which arises
due to the guiding center drift. To show it we can calculate the
corresponding contribution to the diffusion coefficient $ \delta D.$
Following [8,9] we write: 
         \begin{equation}
\delta D = \tau \frac{g}{2 \pi} \int \limits_{- \pi /g}^{\pi /g} V_x^2 (Y)
d Y.
                               \end{equation} 
 This term $ \delta D $ gives the additional contribution to the $ ''x''$
component of the diffusion tensor $ D. $ The latter is connected with the
CF conductivity through the Einstein relation $ \sigma_{\alpha
\beta}^{cf} = N e^2 D_{\alpha \beta} \; (N $ is the CF density of states).
Substituting Eq.(11) into this relation we obtain the expression for this
diffusion correction to $ \sigma_{xx}^{cf} $ which coincides with the last
term in Eq.(8).

According to the HLR theory, the 2DEG resistivity tensor $ \rho $ equals:
$ \rho = \rho^{cf} + \rho^{cs} $ where $ \rho^{cf} $ is the CF resistivity
tensor $ \bigg (\rho^{cf} = (\sigma^{cf})^{-1} \bigg) $ and the
contribution $ \rho^{cs} $ arises due to a fictitious magnetic field which
originates from the Chern-Simons formulation of the theory. The latter has
only off diagonal elements. Hence the diagonal components of the 2 DEG
resistivity tensor coincide with the corresponding components of the CF
resistivity tensor $ \rho^{cf}. $ After straightforward calculations we
arrive at the result:
         \begin{equation}
\rho_{||} =
\rho_{xx} \approx \frac{1}{\sigma_0} \left \{1 + \left (\frac{\Delta
B}{B_{eff}} \right )^2 \chi_1 (gR) \right \}^{-1};
                               \end{equation} 
         \begin{equation}
\rho_\perp =
\rho_{yy} \approx \frac{1}{\sigma_0} \left \{1 + \left(\frac{\Delta n}{n}
k_F l \right )^2 \frac{\chi_2 (gR)}{1 + (\Delta B/B_{eff})^2 \chi_3 (gR)}
\right \};
                               \end{equation} 
 Here $ \chi_i (gR) = \alpha_i J_0^2 (gR) + J_1^2 (gR) \; (i = 1,2,3) $
and the coefficients $ \alpha_i $ are given by the expressions:
         \begin{equation}
\alpha_2 = 0;
\qquad \quad
\alpha_1 = - \frac{1}{2} \,\frac{(\Omega \tau)^2}{1 + (\Omega \tau)^2};
\qquad \quad
\alpha_3 =  \frac{(\Omega \tau)^2}{1 + (\Omega \tau)^2}.
                               \end{equation} 

When the density modulation is very weak $\displaystyle{\bigg(\frac{\Delta
n}{n} k_F l << 1 \bigg)}$ the corrections to the magnetoresistivity are
small and we can neglect them. Under this condition the inhomogeneity of
the effective magnetic field does not significantly affect dc transport. 
For stronger modulation $\displaystyle{\bigg(\frac{\Delta n}{n} k_F l \sim
1 \bigg)}$ the resistivity component $ \rho_{||}, $ as previously, depends
weakly on modulations but $ \rho_{\perp} $ is significantly changed.

We can easily extend our consideration to include higher harmonics of the
periodic density modulation. Suppose that the correction $ \Delta n {\bf
(r)} $ has the form: 
         \begin{equation}
\Delta n {\bf (r)} = 
\sum \limits_{s=1}^\infty \Delta n_s \sin (s g y)
                               \end{equation} 
 In this case we have: 
         \begin{equation}
B {\bf (r)} = B_{eff} \bigg \{1 + \sum \limits_{s=1}^\infty \beta_s \sin
[s g Y - s g \cos (\Omega \tau)] \bigg \}
                               \end{equation} 
 where $ \beta_s = \Delta B_s / B_{eff} $ and $ \Delta B_s $ is the
Fourier transform of the correction $ \Delta B \bf (r). $ For weak
modulations we can assume $ \beta_s << 1. $ Proceeding in a similar way as
before we arrive at the following results for the correction $ \Delta v_x
$ and the cyclotron frequency $ \Omega $ averaged over the CF cyclotron
orbit: 
         \begin{equation}
V_x (Y) = v_F \sum \limits_{s=1}^\infty \beta_s \cos s g Y J_1 (s g R);
                               \end{equation} 

         \begin{equation}
\Omega (Y) = \Omega \bigg [1 + \sum \limits_{s=1}^\infty \beta_s 
\sin (sgY) J_0(sgR) \bigg ].
                               \end{equation} 
 As before $ V_y (Y) = 0.$

Using these expressions we can obtain the following results for the
desired resistivity components
         \begin{equation}
\rho_{||} =
\rho_{xx} \approx \frac{1}{\sigma_0}
\left [1 + \sum \limits_{s=1}^\infty \beta_s^2 \chi_1
(s g R) \right ]^{-1}; 
                               \end{equation} 
         \begin{equation}
\rho_{\perp} =
\rho_{yy} \approx \frac{1}{\sigma_0} \left [1 + 
\frac{\displaystyle
(\Omega \tau)^2 
\sum \limits_{s=1}^\infty \beta_s^2 \chi_2 (s gR)
}{\displaystyle
1 + \sum \limits_{s=1}^\infty \beta_s^2 \chi_3 (s gR)} \right ].
                               \end{equation} 
 Fos a nearly harmonic density modulation we can neglect all coefficients
$ \beta_s \; (s \ne 1) $ and suppose that $ \beta_1 \approx \Delta B /
B_{eff}. $ As a result Eqs.(19),(20) turn into (12),(13). Now we turn to
consider different behavior of the components $ \rho_{\perp} $ and $
\rho_{||} $ as functions of magnetic field. First we note that all the
considerations up to this point are based on the assumption that the
CF--FS is a circle. 

 The magnetic field dependence of the magnetoresistivity is determined by
the parameter $ gR. $ For $ gR \sim 1 $ the functions $ \chi_i (s g R )$
at $ s = 1 $ take values of the order of unity and can increase upon
increase of $ B_{eff}.$ For $ \displaystyle{\frac{\Delta n}{n}k_F l \sim
1, \; \frac{\Delta B}{B_{eff} } << 1} $ the features of dependence of
magnetoresistivity $ \rho_{\perp} $ on the magnetic field is determined by
the function $ \chi_2 (gR)\; (s = 1) $ in the numerator of Eq.(20). The
increase of $ \chi_2 (gR) $ upon increase of the effective magnetic field
corresponds to a minimum in the magnetic field dependence of $
\rho_{\perp} $ around $ \nu = 1/2.$ Such a minimum was observed in the
experiments [1,10]. The magnetic field dependence of the resistivity
component $ \rho_{||} $ is determined by the first term of the sum $
\displaystyle{\sum \limits_{s =1}^\infty \beta_s^2 \chi_1 (s q R)} $ in
the denominator of Eq.(19). Within a certain range of magnitude of the
field $ B_{eff} $ this term may increase on increase of $ B_{eff} $ and
thus give a maximum in the dependence of this resistivity component on the
effective magnetic field at about $ \nu = 1/2. $ The maximum can be
developed when the CF mean free path and the modulation period are large
enough to obey the inequalities $ l > R $ and $ gR \sim 1 $ for
sufficiently small $ B_{eff}. $ However this maximum is too small in
magnitude (because of the small factor $ \beta^2)  $ and narrow compared
with a maximum in the magnetic field dependence actually observed in
experiments [1]. We conclude therefore that the maximum in the magnetic
field dependence of the resistivity $ \rho_{||} $ cannot be succesfully
described within the framework of the same approach which gives rather
good results for the component $ \rho_{\perp}. $

 We conjecture, that this discrepancy mainly originates from a noncircular
shape of the CF-FS which was not taken into account in the original HLR
theory [2,3], and also not considered here in deriving the formulas
(19),(20) for the resistivity components. It follows from these
expressions (19),(20) that in the absence of density modulations both
resistivity components $ \rho_{||} $ and $ \rho_{\perp} $ do not depend on
the magnetic field. This is correct only for a circular CF--FS.  However
the CF--FS may have a noncircular shape due to the effect of the
crystalline field of adjacent layers of GaAs/AlGaAs.  For a modulated 2DEG
the CF--FS can undergo an extra distortion because of the modulation
itself [5]. For a noncircular geometry of the CF--FS the resistivity
components do depend on $ B_{eff} $ even for unmodulated 2DEG, as we now
show.

Consider an unmodulated CF system with a closed FS of arbitrary shape. It
follows from Eq.(7) that at $ \nu = 1/2 \; (B_{eff} = 0)$ we have $
\rho_{xx} (\nu = 1/2) = 1/\sigma_0 a_{xx }, \;  \rho_{yy} (\nu = 1/2) =
1/\sigma_0 a_{yy}.$ The dimensionless coefficients $a_{\alpha \beta} $
equal:
         \begin{equation}
a_{\alpha \beta} = \sum \limits_k w_{-k}^\alpha w_k^\beta;
                               \end{equation} 
 $ w_k^\alpha = v_k^\alpha m_c \sqrt{2 \pi/ A;} \;\; A $ is the area
enclosed by the CF--FS.

Away from $ \nu = 1/2 $ when $ B_{eff} $ is large enough to satisfy an
inequality $ \Omega \tau >> 1 $ we obtain: $ \sigma_{xx}^{cf} = \sigma_0
a_{xx}/(\Omega \tau)^2; \; \sigma_{yy}^{cf} = \sigma_0 a_{yy}/(\Omega
\tau)^2;  \; \sigma_{xy}^{cf} = -\sigma_{xy}^{cf} = \sigma_0 a_{xy}/\Omega
\tau. $ This gives the following approximations for resistivity
components:  $ \rho_{xx} = \rho_{xx} (\nu = 1/2)/ b; \\ \rho_{yy} =
\rho_{yy} (\nu = 1/2)/ b $ where $ b = a_{xy}^2 / a_{xx} a_{yy}. $

The value of this parameter $ b $ is determined from the geometry of the
CF--FS. For a circular CF--FS we have $ a_{xx} a_{yy} = a_{xy}^2 \; (b =
1). $ Then it can be shown that the asymptotics for resistivity components
at the limits of zero and strong effective magnetic field coincide with
each other (resistivity components $ \rho_{xx} $ and $ \rho_{yy} $ do not
depend on $ B_{eff} ). $ In the general case, however, $a_{xx} a_{yy} \ne
a_{xy}^2 \; (b \ne 1) $ and asymptotically the resistivity components at
the limits of zero and strong $ B_{eff} $ can differ; i.e.  they depend on
the magnetic field. When $ b < 1 $ the resistivity components at strong
effective magnetic field take on values greater than at $ B_{eff} =0 \; 
(\nu = 1/2), $ which agrees with the assumption that magnetoresistivity
increases upon increase of the field $ B_{eff}. $ For $ b > 1 \; \rho_{xx}
$ and $ \rho_{yy} $ in the limit of strong magnetic field $ B_{eff} $ are
smaller than at $ \nu = 1/2 $ which can correspond to a maximum around $
\nu = 1/2 $ in the dependence of both resistivity components on $ B_{eff}. 
$

It follows from our results (19),(20) that the inhomogeneous magnetic
field arising due to the density modulations influences the resistivity $
\rho_{\perp} $ significantly more than the resistivity $ \rho_{||}. $ The
effect of the inhomogeneous magnetic field can predominate for $
\rho_{\perp} $ whereas the magnetic field dependence of $ \rho_{||} $ is
determined predominantly by the CF--FS geometry. As a result: a minimum of
the resistivity $ \rho_{\perp} $ at $ \nu = 1/2 $ arising due to the
modulations of the effective magnetic field can match a maximum of the
resistivity $ \rho_{||} $ originating from the geometry of the CF--FS. We
suppose that such a maximum was observed in the experiments [1] in the
magnetic field dependence of the resistivity $\rho_{||}.$

 We also remark that the resistivity $ \rho_{||}$ can be strongly
influenced by the mobility modulations which occur due to the presence of
the modulating field whereas $ \rho_{\perp} $ is nearly independent of
them [7]. Therefore a systematic study of an effect of the CF--FS shape on
the magnetic field dependence of the 2DEG resistivity near $ \nu = 1/2 $
for $ \rho_{||} $ requires that we go beyond the relaxation time
approximation used in this work. We believe, however, that our
semiquantitative estimations capture the essential physics of the effect. 

 The functions $ \chi_i (s g R) $ for $ s > 1 $ near $ \nu = 1/2 $
describe geometrical oscillations of a special kind.  The oscillations
appear due to the commensurability of the CF cyclotron orbit with the
periodic magnetic field induced by the density modulation which
periodically arises upon change (increase or decrease) of $ B_{eff}.$ The
corresponding oscillating structures have to be symmetrically arranged
around $ \nu = 1/2.$

 Similar geometric resonance (so called Weiss oscillations) were observed
in dc transport experiments in both electrostatic and magnetic field
modulated 2DEG in low magnetic fields [11--14]. A theory of these
oscillations in modulated 2DEG systems was first developed within the
framework of a quantum mechanical approach [15--17].  An equivalent
semiclassical approach to the analysis of these phenomena was first
proposed by Beenakker [8] who pointed that Weiss oscillations could be
explained by means of the guiding center drift of cyclotron orbits of the
electrons in the presence of the modulating electric field. The most
complete semiclassical consideration of magneto-transport in a modulated
2DEG is presented in Ref.[7] (See also Ref.[9]).

 The magnetic field dependence of the resistivity component $ \rho_{\perp}
$ is shown in Fig.1. The theoretical curve in Fig.1 is described by
Eq.(20)  where terms corresponding to $ s < 4 $ are kept in the sums over
$ s .$ For simplicity it is assumed that the coefficients $ \beta_s $ in
the retained terms are equal to $ \Delta B / B_{eff} .$ The shape of the
curve is in qualitative agreement with the new experimental data for $
\rho_\perp $ of Willett et al [1]. Two local minima symmetrically arranged
about $ \nu = 1/2 $ correspond to the geometrical Weiss oscillations of
the CFs at the presence of modulating magnetic field.

 Our simplified formula (20) cannot be applied to the region immediately
adjacent to $ \nu = 1/2 $ where the condition $ \Delta B / B_{eff} << 1 $
is not satisfied. So we cannot analyze the effect of channeled orbits of
CFs which occur this region of $\nu $ where $ B_{eff} $ is of the same
order of magnitude as the oscillating correction $ \Delta B. $ To analyze
the dc response of the modulated 2DEG for $ \Delta B \simeq B_{eff} $ we
have to solve the CF transport problem as treated in Ref.[7]. 
Nevertheless our semiquantitative approach gives simple analytical results
applicable for the comparison with experimental data.

In summary, we show that within the single relaxation time approximation,
assuming the CF's have a circular FS we obtain good semiquantitative
agreement with experimental observation of a minimum in transverse
resistivity $ \rho_\perp. $ Assuming the Fermi circle is spontaneously
distorted we can obtain the observed maximum. It is  significant that the
different mechanisms: magnetic field modulation, and Fermi circle
distortion can predominate in transverse and longitudinal geometry.
Additional experiments  can help identify these mechanisms.

We thank Dr.W.R.Willett for kindly giving us his preprint [1] and
Dr.G.M.Zimbovsky for help with the manuscript. Support from a PSC--CUNY
FRAP "In -- Service" Award is acknowledged. 
\vspace{2mm}

-----------------------------------------------------------------
\vspace{2mm}

e-mail: *nzimbov@scisun.sci.ccny.cuny.edu
and **birman@scisun.sci.ccny.cuny.edu.
\vspace{2mm}

1. R.W.Willett, K.W.West and L.N.Pfeiffer ''Observation of Geometric
Resonance of Composite Fermion Cyclotron Orbits with a Fictitious
Magneeetic Field Modulation'', 
Phys. Rev. Lett., (to be published).

2. B.I.Halperin, P.A.Lee and N.Read, Phys. Rev. B {\bf 47}, 7312
(1993).

3. B.I.Halperin, in Low Dimensional Semiconductor Structures, edited by
S.Das Sarma and A.Pinczuk (Wiley, New--York, 1996).

4. An alternative theoretical description of quantum Hall system at $ \nu
= 1/2 $ was recently proposed (See Dung-Hai Lee, Phys. Rev. Lett., {\bf
80}, 4745 (1998);  R.Shankar and Ganpathy Murthy, Phys. Rev.  Lett., {\bf
79}, 4437, (1997);  V.Pasquier and F.D.M.Haldane, cond--mat/ 9712169).  In
the framework of this approach the relevant quasiparticles of the system
are neutral dipolar fermionic objects. However a magneto-transport theory
based on this new physical picture (A.Stern, B.I.Halperin, F. von Oppen,
S.H.Simon, cond--mat/9812135)  gives similar results for observables.

5. N.A.Zimbovskaya and J.L.Birman, Int. J. of Mod. Phys., {\bf 13},  
859 (1999).
 
6. A.D.Mirlin, P.Wolffle, Y.Levinson and O.Entin-Wohlman, Phys. Rev. 
Lett., {\bf 81}, 1070 (1998).

7. Rolf Menne  and R.R.Gerhardts, 
Phys. Rev. B {\bf 57}, 1707 (1998).

8. C.W.J.Beenakker, Phys. Rev. Lett., {\bf 62}, 2020 (1989).

9. R.R.Gerhardts, 
Phys. Rev. B {\bf 53}, 11064 (1996).

10. J.H.Smet, K. von Klitzing, D.Weiss and W.Wegscheider, Phys. Rev.
Lett., {\bf 80}, 4538 (1998).

11. R.R.Gerhardts, D.Weiss and K.von Klitzing,
Phys. Rev. Lett., {\bf 62}, 1173 (1989).

12. R.W.Winkler, J.P.Kotthaus and K.Ploog,
Phys. Rev. Lett., {\bf 62}, 1177 (1989).

13. H.A.Carmona, A.K.Geim, A.Nogaret, P.C.Main, T.J.Foster, M.Henini,
S.P.Beaumont and M.G.Blamire, 
Phys. Rev. Lett., {\bf 74}, 3009 (1995).

14. P.D.Ye, D.Weiss, R.R.Gerhardts, M.Seeger, K. von Klitzing, K.Eberl and
H.Nickel,
Phys. Rev. Lett., {\bf 74}, 3013 (1995).

15. C.Zhang and R.R.Gerhardts,
Phys. Rev. B {\bf 41}, 12850 (1990).

16. F.M.Peeters and P.Vasilopoulos,
Phys. Rev. B {\bf 47}, 1466 (1993).

17. D.Pfannkuche and R.R.Gerhardts, 
Phys. Rev. B {\bf 46}, 12606 (1992).

\vspace{5mm}
\centerline{Figure Caption} 
\vspace{5mm}

Fig.1. dc magnetoresistivity versus the effective magnetic field.
Dashed line --
experiment of Ref.[1], solid line -- theory for the parameters $ n = 1.2
\times 10^{11}$cm$^{-1}$, $\Delta n /n = 0.01$ , $ \lambda = 0.7 \mu m, $
$ l = 10^{-4 }$cm.  The dotted part of the theoretical curve corresponds
to the region of the values of $ B_{eff} $ where Eq.(20)  cannot be
applied.

\begin{figure}[htbp]
 \includegraphics[width=0.750 \textwidth]{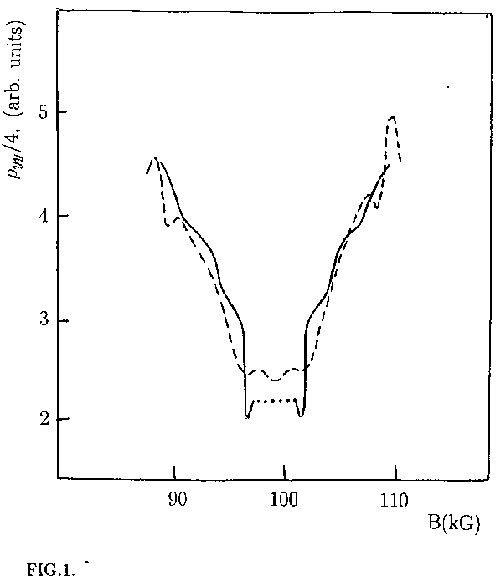}
\end{figure}


\newpage

\end{document}